# Academic match-makers in sociology: Their role in collaboration network formation

Hongkan Chen[1]   Qingshan Zhou[1*]
Robin Haunschild[2*]   Yi Bu[1*]

1. Department of Information Management Peking University, Beijing 100871

2. Max Planck Institute for Solid State Research, Stuttgart, D-70569

* represents co-corresponding authors

**Abstract:** In modern scientific collaboration networks, certain researchers play a pivotal role in bridging scholars who have never worked together—a phenomenon we term academic "match-makers." Unlike traditional brokers who merely facilitate connections without participating, match-makers actively engage in the resulting collaborations. Despite their potential importance, the prevalence, characteristics, benefits, and long-term trajectory of these individuals remain underexplored. Using the Microsoft Academic Graph (MAG), we operationalized a match-maker as an author who, in a given publication, introduced a first-time collaboration between two co-authors, each of whom had previously collaborated with the match-maker but not with each other. We employed a configuration null model to distinguish observed patterns from random chance. Our findings reveal that the match-maker phenomenon is deliberate, prevalent, and consequential. Among authors with over 20 publications, nearly 30% have served as a match-maker, and the probability of acting as one increased eightfold from 1980 to 2019. Publications involving a match-maker are more likely to appear in high-impact journals and exhibit higher disruptiveness—

[*] Author biographies: Chen Hongkan, male, born in 2001, Ph.D. candidate at the Department of Information Management, Peking University, email: chenhongkan@pku.edu.cn; Robin Haunschild, male, born in 1978, researcher at the Max Planck Institute for Solid State Research, email: R.Haunschild@fkf.mpg.de; Zhou Qingshan, male, born in 1965, professor and doctoral supervisor at the Department of Information Management, Peking University, email: zqs@pku.edu.cn; Bu Yi, male, born in 1994, assistant professor, researcher, and doctoral supervisor at the Department of Information Management, Peking University, email: buyi@pku.edu.cn.

This paper is supported by the National Social Science Foundation of China (#24ZDA078).

particularly in larger teams—suggesting that match-makers help facilitate what we term integrative disruption. Match-makers tend to emerge early in their careers, peaking around the 20th publication and at an academic age of roughly ten years. While nearly all match-makers eventually experience "abandonment" in the sense that the connected researchers later collaborate without them, their continued involvement remains substantial and is driven by research needs rather than structural factors. This reframes abandonment not as exclusion but as a natural evolution within project-based collaborations. The academic match-maker phenomenon is a strategic feature of collaboration networks characterized by early-career emergence, context-dependent persistence, and tangible contributions to high-impact, disruptive research. These findings deepen our understanding of the microdynamics that shape scientific teamwork and knowledge integration, with implications for theories of structural holes, network orchestration, and the sociology of science.

# Introduction

In the modern academic landscape, the formation of collaborative networks is profoundly shaped by the interconnections and interactions among researchers. With the rapid development of technology and information dissemination, the patterns of academic collaboration have undergone significant changes, with cross-disciplinary and cross-field collaboration increasingly becoming a primary driver of academic progress. Within these increasingly complex academic collaboration networks, some researchers not only participate directly in collaborations but also act as bridges connecting different scholars and research teams; these individuals are referred to as academic "match-makers". This is also called the *tertius iungens* event (Obstfeld, 2005). By linking different researchers, academic match-makers help facilitate cross-disciplinary collaboration, thereby promoting the dissemination of knowledge and the generation of innovation. Beyond cross-disciplinary bridging, the role of match-makers is equally vital within disciplines. By linking researchers whose capabilities are suitably complementary, they catalyze the shift from solitary endeavors to collaborative research. In doing so, they unlock a synergy that leads to far more significant academic outcomes than individuals could achieve alone.

The role of academic match-makers extends far beyond that of mere collaboration facilitators. By leveraging their resources and relationships within the academic network, they often provide unique value in collaborations. These match-makers are sometimes called "brokers" or "brokerages." They not only help to establish connections between different scholars but may also influence the direction of academic research and the integration of fields. We deliberately avoid the term "broker"/ "brokerage" because a broker, by definition, does not typically participate in the collaboration itself, which contradicts the essence of scientific research. We believe the term "match-maker" is more appropriate, as it describes someone who matches individuals with diverse capabilities across multiple dimensions to initiate collaborations. However, despite the important role academic match-makers play in collaboration networks, existing research has paid limited attention to members' match-making behavior, while focusing primarily on other forms of brokerage events (Collet et al., 2014; Garcia-Morante et al., 2025; Jessani et al., 2016, 2016; Lam, 2018; Lomas, 2007). Guimerà et al. (2005) and Bachmann et al. (2026) have examined such match-making behavior at the team level and individual level, respectively. The characteristics, mechanisms of action, and influence of academic match-makers within academic collaboration networks have not yet been fully explored.

Research on the phenomenon of academic match-makers is highly significant. Firstly, understanding the role of academic match-makers helps to reveal the dynamic mechanisms of academic collaboration, particularly the formation and development processes of cross-disciplinary collaboration. Secondly, academic match-makers may play a dual role: While they can facilitate the establishment of collaborations, they might also be gradually excluded or "abandoned" in subsequent collaborations as their role potentially diminishes. The persistence of their role and the phenomenon of being "abandoned" require further investigation, especially as collaborations mature and their influence might wane. Finally, the "benefits" reaped by academic match-makers also warrant in-depth study. By analyzing how many additional collaborative opportunities academic match-makers can generate and their influence on other scholars within the academic field, we can better understand their value in the academic collaboration network.

To this end, several critical research questions warrant further investigation to delineate the full scope of the match-maker phenomenon. First, it is essential to determine the prevalence of this phenomenon (RQ1): Is the role of a match-maker a rare occurrence or a common structural position within collaboration networks? Second, we aim to examine the outcomes associated with their bridging activity (RQ2): Does the involvement of a match-maker generate tangible benefits, such as enhancing the citation impact of resulting publications or facilitating the network expansion of the connected authors? Third, a deeper profile of these individuals is needed (RQ3): What are the defining characteristics of academic match-makers in terms of their career stage, prior productivity, or social capital? Finally, a longitudinal perspective is crucial to understand their trajectory (RQ4): What is the ultimate fate of match-makers—are they subsequently excluded as collaborations mature, or do they continue to play an active role in new research endeavors?

## Materials and methods

### Dataset

We utilized the Microsoft Academic Graph (MAG, 202002 snapshot) as primary data source (Sinha et al., 2015; Wang et al., 2020). The MAG dataset is a widely recognized and comprehensive bibliographic resource, containing over 200 million publications and over 2 billion citation relationships.

MAG employed a bottom-up strategy to construct its field-of-study hierarchy, which comprises 19 major disciplines at its highest level. Our analysis focused specifically on the discipline of sociology. We focused on sociology because it typically features a smaller average number of authors per publication compared to fields like physics or biomedicine. This characteristic minimizes bias when treating author contributions as equal in empirical analyses, ensuring that the match-makers we identified genuinely represent collaborative links between individuals rather than artifacts of hyper-authorship. To define our target population of authors, we first identified all authors who have authored at least one publication classified under sociology. Subsequently, all journal and conference publications by these authors—regardless of sociology—were included in our subsequent analysis. The final dataset comprised 18,030,778 author-publication relationships between 2,749,999 authors with 13,286,019 publications, with the distribution of publications by year shown in Figure S1. Additionally, we used journal ISSN, eISSN, and names to match against the JCR 2023 journal quartiles; about 75% (15,385/20,648) journals were matched from JCR. The indicators introduced later were obtained from SciSciNet-v2 (Lin et al., 2023).

### Indicators

To assess the academic impact of match-maker publications, we employed two measures. The first is journal quality, operationalized as whether a publication is in a Q1 journal according to the JCR 2023 ranking. The second is citation performance, measured by cumulative citation counts at three, five, and ten years after publication (C3, C5, and C10, respectively). To eliminate the confounding effects of publication year and team size on citation accumulation, citation rankings were normalized within the same year and the same team size. Publications in the top 10% of this normalized distribution were identified as high-impact.

For disruption, we adopted the disruption index (DI) proposed by Funk and Owen-Smith (2017), which measures the extent to which a study renders its prior work less cited by subsequent research. Following Chen et.al (2025), only publications with at least five references and five citations are included in the disruption analysis. Given that DI is directly related to reference count and that large teams tend to consolidate innovations originating from small teams, rankings are normalized within groups defined by the same year, the same team size, and the same reference count range. Since DI ranges from –1 (least disruptive) to 1 (most disruptive), we consider two complementary indicators: the proportion of publications with DI ranked in the top 10%, and—using the intuitive threshold of zero—the proportion with DI greater than 0 (indicating positive disruption). These together provide a more comprehensive assessment of disruptiveness across publication sets.

Novelty is measured based on the combinatorial novelty of references. Following Uzzi et al. (2013), we assess the degree to which a publication combines journal pairs that have rarely or never appeared together in previous works. As with disruption, novelty rankings are normalized within groups defined by publication year, team size, and reference count range. We report the proportion of publications with novelty index ranked in the lowest 10% within their group (i.e., the top 10% in terms of novelty), as well as the proportion with novelty index less than 0—both of which reflect a tendency toward atypical knowledge combinations.

## Definition of match-makers

In this study, we operationalized the concept of a "match-maker" as an individual who acts as a bridge in academic collaborations, similar with the definition of the study of Bachmann et al.(2026). Specifically, a scholar was identified as a match-maker, denoted as *a*, within the context of a specific academic publication if they facilitated the first collaborative tie between two other researchers, *b* and *c*, who had no prior co-authorship history but had prior co-authorship history with *a*.

The identification procedure adheres to a specific temporal sequence involving a match-maker (*a*) and two other researchers (*b* and *c*): (1) At time t, a publication co-authored by *a*, *b*, and *c* appears, marking the first recorded collaboration between *b* and *c*. (2) Prior to time t, an established co-authorship relationship exists between *a* and *b*. (3) Prior to the same time t, *an* established co-authorship relationship also exists between *a* and *c*. (4) Prior to time t, an established co-authorship relationship between b and c does not exist. When these conditions are met, a is operationally defined as the match-maker for the pair b and c in that specific publication. We defined the researcher with more co-publications with the match-maker as b. Figure 1A illustrates a true sample for better understanding. In total, we identified 80,923 match-makers and 164,144 connected researchers across 160,970 publications. To ensure the validity of our findings, we subsequently restricted our analysis to publications involving only one match-maker. This yielded a refined sample of 56,066 match-makers and 137,646 connected researchers from 114,526 publications.

To quantify the phenomenon of "abandonment," we examined the subsequent collaboration patterns after the first co-authorship publication. We measured whether the match-maker a is gradually excluded from the ongoing collaboration between b and c. This was assessed by comparing two counts derived from the publication record: (1) $N_{abc}$ represents the number of subsequent publications co-authored by all three individuals (a, b, and c) after their first co-authorship publication. $N_{bc}$ represents the number of subsequent publications co-authored by b and c without a.

The operational criterion for abandonment was defined as $N_{bc} > N_{abc}$. This inequality indicated that the collaboration between b and c continued and intensified independently, significantly surpassing their continued collaboration with the original match-maker a. This metric served as a proxy for the match-maker's diminishing influence and their eventual exclusion from the collaboration network they initiated. To mitigate potential biases arising from a truncated time window, we restricted this analysis to match-making events that occurred in or before 2015.

## Configuration null model

To ensure that our observed results are not artifacts of random chance, we employed a configuration null model (CNM) to test the statistical significance of our findings (Bollobás, 1998). The core principle of this model is to randomize the author-publication associations within the academic network while preserving key structural properties. Specifically, we implemented a constrained randomization procedure grouped by field of study and publication year. As illustrated in

Figure S2, within each group, we shuffled the links between authors and publications, under the constraint that the number of authors per publication and the number of publications per author remained unchanged from the original data. This process generated randomized versions of the collaboration network that retained the basic productivity and collaboration size patterns but destroyed the higher-order structure formed by strategic match-making.

# Results

## Overview of the match-maker phenomenon

Figure 1B indicates that more than 70% of such publications involve only one match-maker, while approximately 23% contain between two and four match-makers. Since it is not feasible to clearly distinguish or interpret which specific match-maker played the decisive role in publications with multiple match-makers, we have chosen to focus our subsequent analysis exclusively on publications with only one match-maker.

As shown in Figure 1C, the actual data reveal that among authors with more than 20 publications, nearly 30% have served as a match-maker, in contrast to the null model where this probability remains consistently below 10%. As the number of publications increases, the proportion of scientists becoming match-makers rises significantly. Notably, among scientists with over 100 publications, one in two has acted as a match-maker. The subplot of Figure 1C further reveals that, unlike the distribution observed in the null model, the publication counts of scientists who become match-makers in reality exhibit a more concentrated pattern. Specifically, scientists with fewer than 45 publications account for nearly 80% of all match-makers. This finding suggests that in actual academic collaboration, relatively active scholars—particularly those with moderate publication counts—are more likely to play the role of match-makers, facilitating collaboration among different researchers. In summary, while scientists with more publications are more likely to become match-makers, those with fewer publications also play a significant role as academic intermediaries in practice. This indicates that the function of a match-maker is not exclusive to highly prolific authors; under certain conditions, less prolific authors can also foster collaboration through their specialized resources and collaborative networks.

Figure 1D shows a clear upward trend in the rate of being a match-maker, rising from less than 0.005 in 1980 to nearly 0.040 in 2019—an approximately eight-fold increase. This growth suggests that as academic collaboration networks have expanded and scholarly exchange intensified, more researchers have gained opportunities to serve as match-makers, thereby facilitating cross-disciplinary and cross-domain collaboration. The trend may reflect both evolving patterns of scientific collaboration and the growing importance of the match-maker role in enabling such partnerships. As Figure S3 shows, the findings are robust to alternative definitions of active authors.

Figure 1E illustrates the distribution of team sizes in publications involving a match-maker. The results indicate that the vast majority of these teams are relatively small, with nearly

60% consisting of exactly three members and over 30% comprising four to six members. This pattern suggests that the emergence of a match-maker is unlikely to be a random occurrence in large teams but is often an intentional outcome. The subplot of Figure 1E further reinforces this interpretation, highlighting that the role of match-makers is typically a deliberate one, enabling them to play a significant part even within small collaborative teams by fostering connections and facilitating exchanges among different researchers. Smaller team sizes generally allow for more efficient coordination and communication, and match-makers leverage these close-knit environments to bridge gaps and promote the formation and expansion of academic collaborations. Thus, the presence of a match-maker is not random but is closely linked to team structure, collaborative dynamics, and social networks. These findings further validate the strategic and proactive role of match-makers in small-team settings.

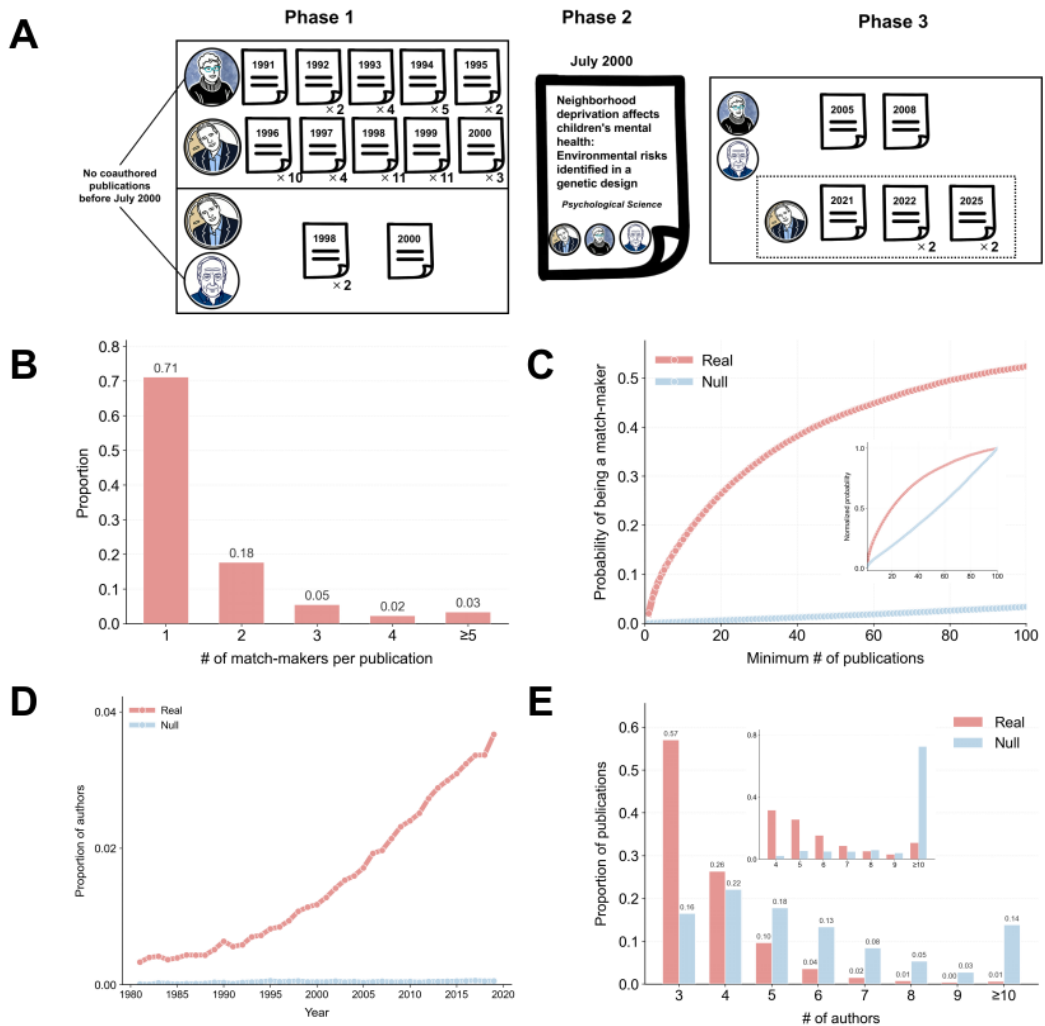

Figure 1. The prevalence of match-makers. (A) The illustration of the match-maker

phenomenon occurred in *Neighborhood deprivation affects children's mental health: Environmental risks identified in a genetic design* (published in July 2000 in *Psychological science*) coauthored by Avshalom Caspi (the match-maker), Alan Taylor (not shown in the figure), Terrie E. Moffitt (*b*), Robert Plomin (*c*). (B) The distribution of the number of match-makers per publication. (C) The probability of an author acting as a match-maker, as a function of their total publication count. The subplot displays the cumulative density distribution of publication counts for all identified match-makers. The dashed line represents the corresponding probability from the null model. (D) The annual proportion of active authors (defined as those who published at least one publication in a given year and had accumulated at least three publications by that year) who served as a match-maker from 1980 to 2019. (E) The distribution of team sizes for publications involving match-makers. The main panel is restricted to publications with only a single match-maker, whereas the subplot displays the distribution for publications containing at least two match-makers.

## The benefit of match-makers

Panels A-C in Figure 2 present the performance of match-maker publications across multiple dimensions, including citation counts, journal quality, disruption, and novelty. The proportion of publications appearing in Q1 journals exceeds 40%, which is notably higher than the baseline proportion of 31.2% observed across all publications. Moreover, we employed propensity score matching (PSM) to construct a control group for each publication, controlling for publication year and average academic age. The results, reported in Figure S3, remain robust, indicating that match-maker publications are indeed more likely to appear in higher-tier journals. In terms of citation impact, the share of match-maker publications ranked in the top 10% does not show a substantial difference from the baseline, though there is a modest tendency toward higher likelihood. Nevertheless, when compared to the PSM-matched control publications, those with a match-maker exhibit a higher citation impact, as shown in Figure S5.

While the proportion of publications with a disruption value greater than 0 does not significantly exceed the overall dataset level, the proportion ranked in the top 10% increases notably with team size—a pattern that runs counter to the finding by Wu et al. (2019) that small teams disrupt, and large teams tend to consolidate prior work. This contrasting trend may suggest that match-makers help large teams achieve genuinely disruptive outcomes.

In terms of novelty, however, a different pattern emerges. The proportion of publications with a novelty index below 0 (i.e., those favoring atypical knowledge combinations) is higher than the overall dataset average. Yet, as team size increases, the proportion ranked in the top 10% of novelty (i.e., the smallest novelty values within each group) declines sharply and even approaches zero. This suggests that match-maker publications do not simply pursue an abundance of unconventional reference combinations. Instead, they appear to achieve disruptive outcomes rooted in traditional combinations, highlighting a more substantive form of scientific advancement.

Figure 2D reveals a positive correlation between the number of match-makers a researcher has encountered and the number of new collaborators gained. On average, however, a single match-maker facilitates connections with fewer than two new collaborators. This implies that for an individual researcher, a match-maker provides access to more than just one new collaborator in the immediate context but does not lead to an excessive number of new connections. This observation is further reflected in Figure 2E, which shows that while a match-maker with more publications has historically helped more researchers, the total number of beneficiaries generally remains between two and four individuals.

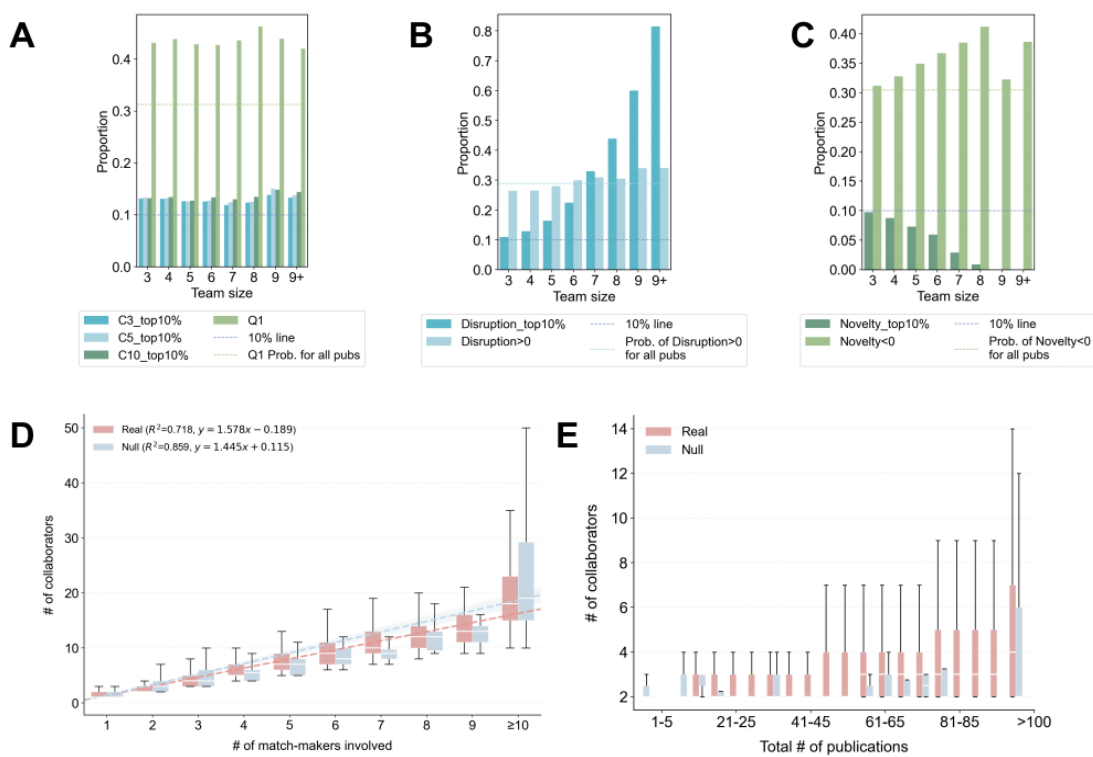

Figure 2. Publication impact and the extent of scientific benefits facilitated by match-makers. (A) For match-maker papers of varying team sizes, the proportion published in JCR Q1 journals and the proportion with citation counts ranked in the top 10%. (B) For match-maker papers of varying team sizes, the proportion with disruption index ranked in the top 10% and the proportion with disruption index greater than 0. (C) For match-maker papers of varying team sizes, the proportion with novelty index ranked in the top 10% and the proportion with novelty index less than 0. (D) The number of distinct collaborators a researcher can access through different number of match-makers. (E) The relationship between the total number of publications by match-makers and the average number of researchers who benefit from them.

## The characteristic of match-makers

Figure 3A indicates that this probability is relatively low when scientists have few publications, increases rapidly as the number of publications grows, and peaks at around 30 publications—with approximately a 0.8% chance of becoming a match-maker. Beyond this point, the probability stabilizes despite a mild downturn, reinforcing that the propensity for match-making emerges early in a researcher's career from the perspective of publications. This result is similar to the finding in physics (Bachmann et al., 2026).

This trend implies that the role of a match-maker resembles an inherent trait that manifests early in an academic career, much like some individuals possess innate social and coordination skills. This may explain why certain researchers exhibit a strong propensity for facilitating collaboration early on-a tendency that remains stable throughout their research trajectory. In other words, becoming a match-maker is not solely a product of later effort, but rather a role that emerges early within the academic collaboration network. This phenomenon can also be analogized to certain inherent aspects of human social behavior: social tendencies are often intrinsic and enduring, rarely changing completely within short time frames. Thus, whether one becomes a match-maker and the role they play in collaborations may be closely tied to individual dispositions and early academic collaboration experiences. Furthermore, the necessity of additional knowledge or abilities for a planned project might cause researchers to become match-makers.

Figure 3B reveals that the peak occurrence of match-makers in the real network centers around an academic age of ten years, whereas the peak in the null model appears at 25 years. This finding further demonstrates that becoming a match-maker is not the exclusive privilege of senior scientists. In reality, most researchers exhibit the potential to act as match-makers early in their careers, contributing to the formation and expansion of academic networks through their role in fostering collaborations.

Figure 3C shows a roughly linear relationship, indicating that becoming a match-maker generally correlates with a higher publication count; in other words, those who become match-makers later in their careers tend to have accumulated more publications. However, it is noteworthy that the majority of match-making events occurs early in a researcher's career, particularly within the first ten years of academic age or before reaching 15 publications. This suggests that while a positive correlation exists between publication count and the likelihood of becoming a match-maker, the potential for such a role often manifests at an early stage.

Figure 3D shows that in the majority of cases, the number of co-authored publications with each collaborator is up to 20. Even when the match-maker has a high number of publications with one collaborator, the other collaborator tends to have a relatively low co-authorship count, reflecting a typical "experienced researcher guiding a newcomer" collaboration pattern. The subplot of Figure 3D further shows that as the number of co-authored papers with the more frequent collaborator increases, the number of co-authored papers with the less frequent collaborator also rises, albeit disproportionately — by approximately a factor of ten. This suggests that when facilitating collaborations, match-makers do take into account the collaborative potential among different partners, even if the resulting collaboration volumes are uneven.

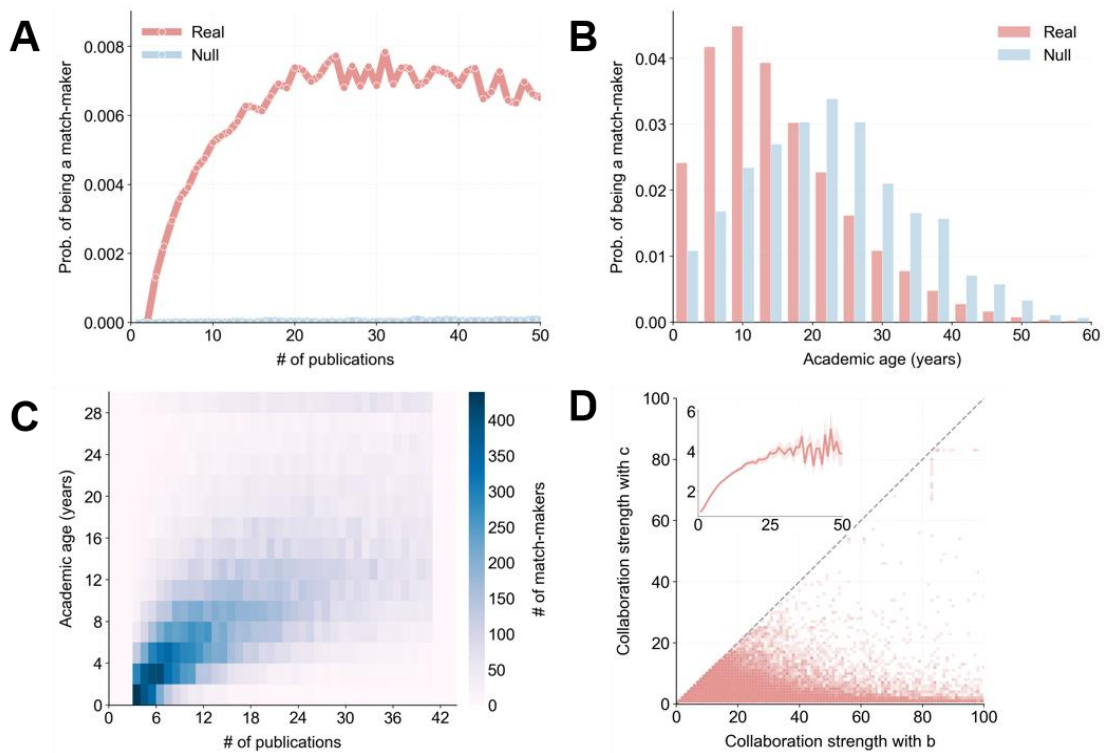

Figure 3. Match-makers tend to emerge early in a scientific career. (A) The probability of a scientist becoming a match-maker as a function of their publication sequence. (B) The distribution of academic age when scientists first act as match-makers. (C) Relationship between publication count and academic age at the time of first match-making. (D) Collaboration frequency between the match-maker and each of the two researchers they connect. The subplot shows the association between collaboration frequency with the more versus less frequent collaborator.

To mitigate the potential influence of informal mentorship, we conducted robustness checks by excluding samples where the minimum academic age of researchers b or c was five years or less. The results remained robust, as shown in Figure S6. In addition, when we imposed a stricter criterion requiring that the match-maker had previously published at least three publications with each of the two researchers, the findings also held (see Figure S7).

## The future of match-makers

Figure 4A reveals that, in contrast to the null model where almost no match-makers are "abandoned" by their collaborators, nearly all match-makers experience abandonment in the real world. Figure 4B and Figure 4C show that although abandonment is a common experience, its actual rate is significantly lower than that in the null model. Furthermore, as the number of co-authored papers between the two connected collaborators increases, the rate of abandonment also rises. Figure 4D indicates that, regardless of the subsequent collaboration frequency between the two collaborators, the first abandonment of the match-maker generally occurs about two years after the first collaboration. Figure 4E demonstrates that the proportion of match-makers being abandoned remains relatively stable, at around 0.5, irrespective of the career stage at which they facilitated the collaboration.

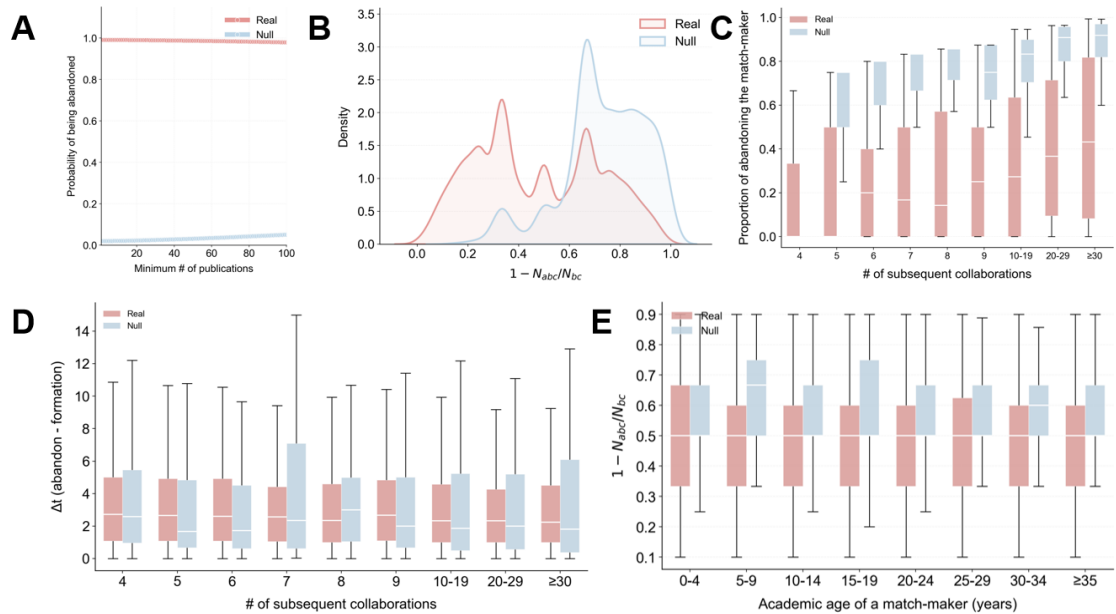

Figure 4. Characteristics of match-maker abandonment. (A) Abandonment rate of match-makers, stratified by their total publication count. (B) Proportion of subsequent collaborations from which the match-maker is excluded (analysis restricted to cases with at least three subsequent co-authored publications to minimize outlier effects). (C) Relationship between match-maker abandonment rate and the collaboration frequency between the connected researchers. (D) Time to first abandonment of the match-maker, categorized by the subsequent collaboration intensity between the connected collaborators. (E) Abandonment proportion of match-makers across collaborations initiated at different career stages.

## Discussion

We operationalized and empirically examined the concept of the academic "match-maker," a researcher who actively facilitates and participates in the first collaboration between two previously unconnected scholars. Our findings, grounded in the publication records of authors publishing in sociology and rigorously tested against a configuration null model, reveal that this phenomenon is a deliberate, prevalent, and consequential feature of scientific collaboration networks. The evidence suggests that match-making is not a random artifact of publication volume but a strategic behavior associated with enhanced research impact, characterized by early-career emergence, and defined by a complex, context-dependent form of persistence despite near-universal structural "abandonment." These results contribute to a deeper understanding of the micro-dynamics that shape collaborative science, resonating with and extending several key theoretical frameworks in the sociology of science and social network analysis.

The widespread and increasing prevalence of match-makers underscores their integral role in the modern research enterprise. The eightfold increase in the propensity for this role since the 1980s aligns with broader shifts towards team-based and interdisciplinary science (Larivière & Gingras, 2010; Wuchty et al., 2007). That a significant portion of active scholars serve as match-makers at least once, and that the phenomenon is concentrated in small to mid-sized teams, challenges the notion that brokerage is exclusive to a small elite or large, amorphous groups. Instead, it points to a distributed model of network orchestration where many researchers, at various productivity levels, actively engage in bridge-building. This finding enriches our understanding of how collaborative networks evolve beyond mere preferential attachment or homophily (Zhang et al., 2018), highlighting intentional triadic closure as a common mechanism (Rivera et al., 2010). The stark contrast with the null model confirms that this activity is non-random and strategically embedded within the social fabric of academia.

The tangible benefits associated with match-maker-facilitated publications, namely, a higher likelihood of appearing in top-tier journals and a tendency toward greater disruptive impact, provide compelling evidence for the value of this integrative role. This aligns with the theory of "structural holes" (Burt, 1995) where actors who bridge disconnected network clusters can access diverse information and synthesize it for innovative advantage (Burt, 2004). Our match-maker embodies a specific, participatory form of such brokerage. Crucially, our nuanced findings on disruption and novelty suggest that the match-maker's key contribution may not be in forging entirely unprecedented knowledge combinations (Uzzi et al., 2013) but in enabling what we might term integrative disruption. They appear to help teams, particularly larger ones, synthesize established but previously separate knowledge streams in novel ways that challenge existing paradigms. This offers a potential explanation for the counter-intuitive capacity of some larger teams involving a match-maker to achieve disruptiveness, contrasting with the general trend of consolidation in big teams (Wu et al., 2019). The match-maker thus acts as a social and cognitive linchpin, facilitating the recombination necessary for transformative work.

A pivotal finding is the distinct career-stage signature of the match-maker role. Its peak probability around the 20$^{th}$ publication and an academic age of ten years suggests it is predominantly an early- to mid-career activity. This timing may reflect a period of maximum network expansion and project diversification, where researchers actively seek complementary expertise to advance their work (Leahey et al., 2017). It challenges the assumption that network influence and brokerage are primarily functions of seniority and accumulated status. Instead, it posits match-making as a dynamic, project-driven strategy employed by researchers building their reputations and portfolios. The bounded influence of a single match-maker connecting an individual to only one or two new collaborators on average further underscores the selective and effort-intensive nature of this role, constrained by the cognitive and temporal limits of maintaining productive ties (Dunbar, 1992).

Perhaps the most theoretically significant results concern the longitudinal fate of match-makers. The near-universal experience of "abandonment," defined by the bridged pair subsequently collaborating more without the match-maker, is, at a structural level, consistent with theories of network efficiency and triadic closure. Once a direct tie is established, the intermediary's structural necessity diminishes (Burt, 2005). However, a deeper analysis reveals a more substantive story. The fact that the match-maker's actual rate of continued exclusion is far lower than in randomized networks, and that the timing and likelihood of "abandonment" are not tied to their career stage, suggests their role is *not* merely structural but contingent on research needs. This pattern indicates that match-makers are not passive brokers who are inevitably discarded but active participants whose involvement waxes and wanes with the specific requirements of a research trajectory. Their initial role is catalytic and integrative, yielding a high-impact publication. Their subsequent participation is then governed by the evolving intellectual demands of the collaboration, rather than by a deterministic network logic. This reframes "abandonment" not as exclusion but as a natural evolution in a project-based ecosystem, reflecting a form of Mertonian "organized skepticism" and communal labor where roles are fluid (Merton, 1973).

Several limitations of this study must be acknowledged. First, its focus on sociology, while methodologically sound for controlling team size effects, may limit generalizability to fields with vastly different collaboration norms, such as high-energy physics or biomedicine (Jones et al., 2008). Second, our co-authorship-based operationalization necessarily captures only formalized outcomes of match-making, missing the rich substratum of informal introductions, recommendations, and networking that may not lead to immediate publication. Third, while we infer intentionality from patterns, we lack qualitative data on the motivations, negotiations, and power dynamics within these triads. Fourth, the analysis of "abandonment" is behavioral; we cannot discern whether a match-maker's non-participation is a voluntary choice, a natural conclusion, or an outcome of exclusion. Future research should build upon these foundations. Cross-disciplinary comparative studies, for instance, are crucial to test the universality and field-specific variations of the match-maker phenomenon. Qualitative investigations, including interviews and ethnographies, are needed to open the "black box" of how match-making is initiated and negotiated. Research should also explore demographic, institutional, and geographic dimensions to assess potential biases in who gets to act as a match-maker and for whom. Furthermore, investigating the long-term career impact of both being a match-maker and being connected by one would illuminate the full personal and collective returns on this form of social and intellectual brokerage.

# Supplementary Information

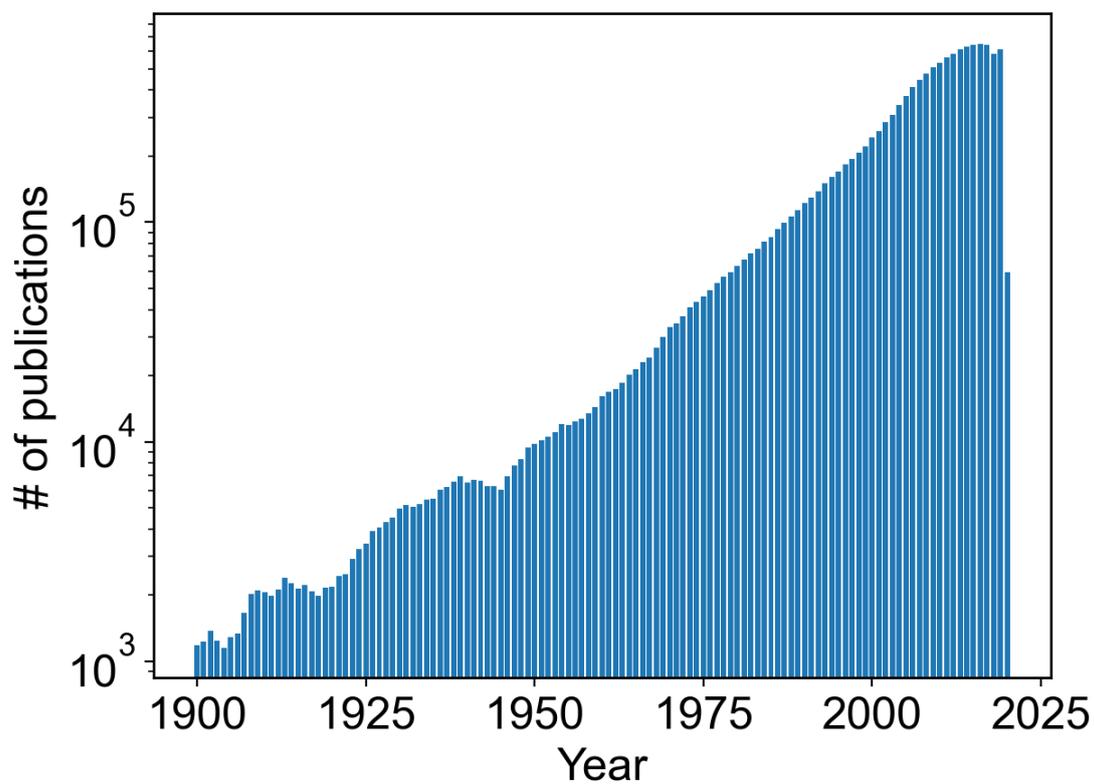

Figure S1. Distribution of publication years since 1900.

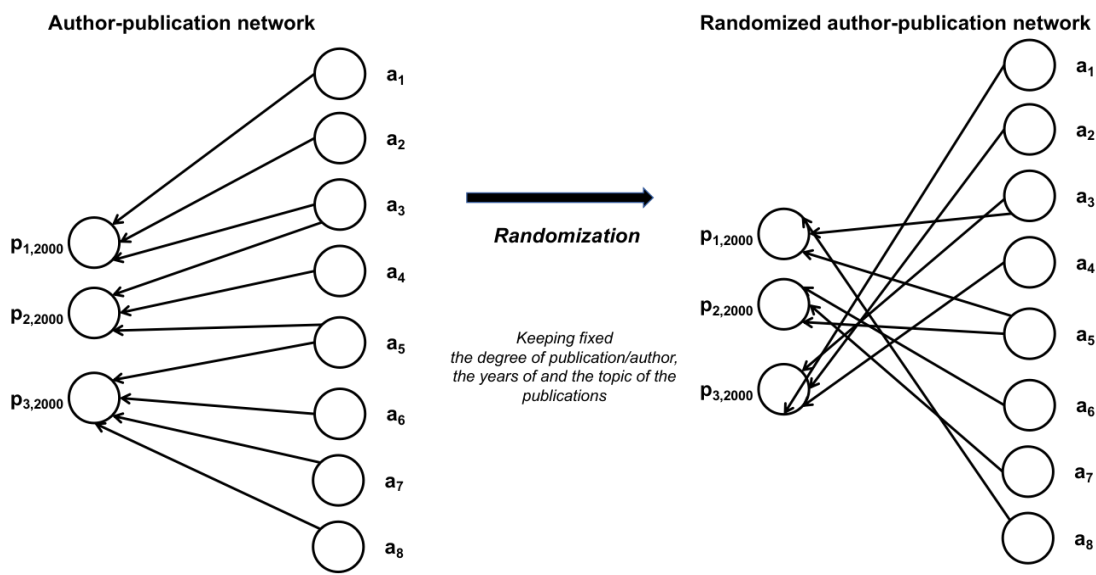

Figure S2. Configuration Null Model. The randomized author-publication network (right-hand side) is obtained by applying a CNM to the observed author-publication network (left-hand

side). The randomization is performed fixing the years of and the topic of the publications.

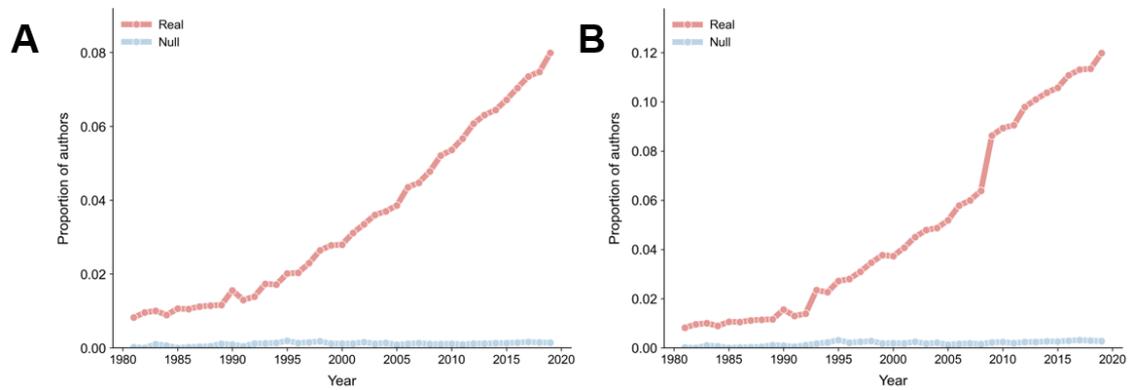

Figure S3. The annual proportion of active authors (defined as those who published at least one publication in a given year and had accumulated at least three publications by that year) who served as a match-maker from 1980 to 2019. (A) Active authors are those who published at least three publications in a given year. (B) Active authors are those with an annual publication count above or equal to the 90-percentile threshold for that year (yearly publication count thresholds are three before 1992, four before 2008, and five untill 2019).

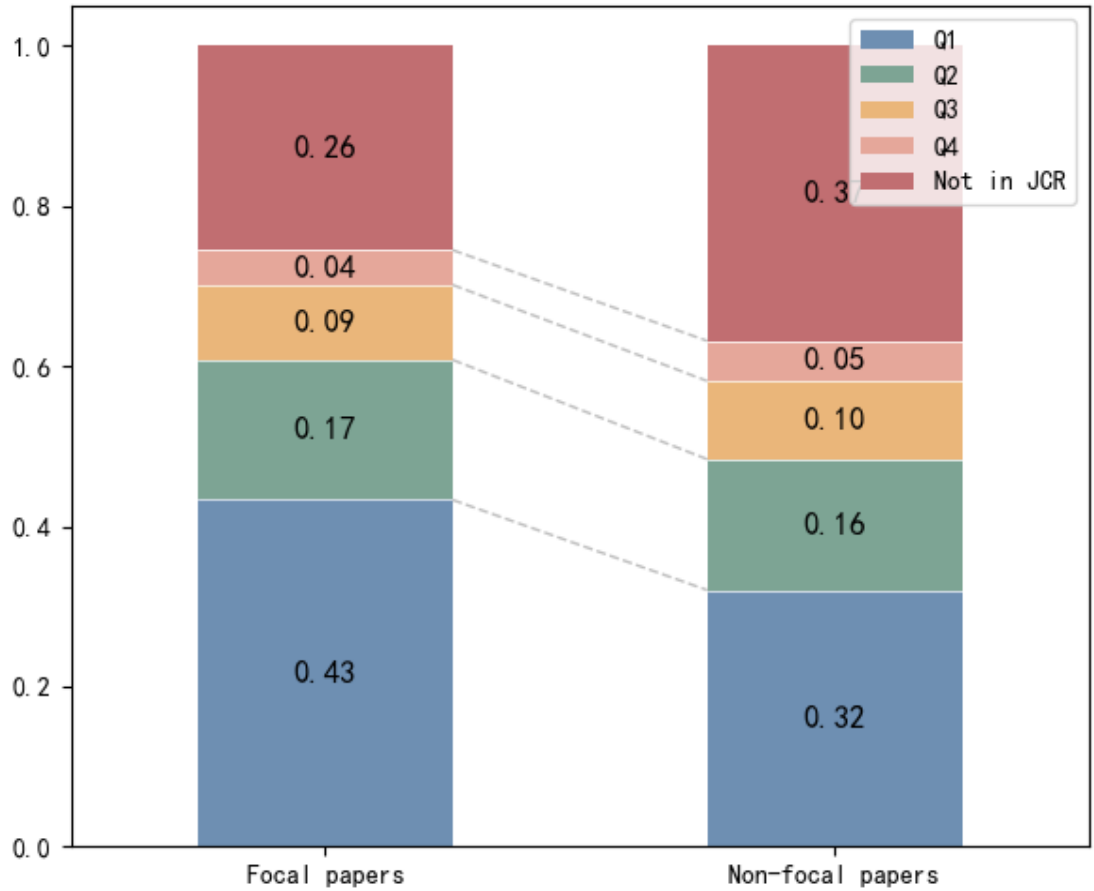

Figure S4. Journal quartile distribution (based on JCR 2023) compared to that of PSM-matched publications, controlled for publication year and average academic age.

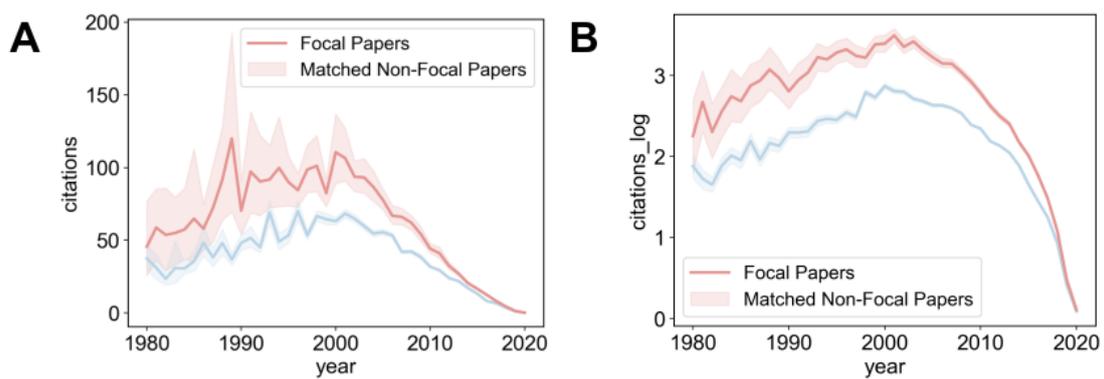

Figure S5. Temporal evolution of the average citation impact for focal publications compared with PSM-matched publications, controlled for publication year and average academic age. Panels A and B present the results for raw and log-transformed citation impact, respectively.

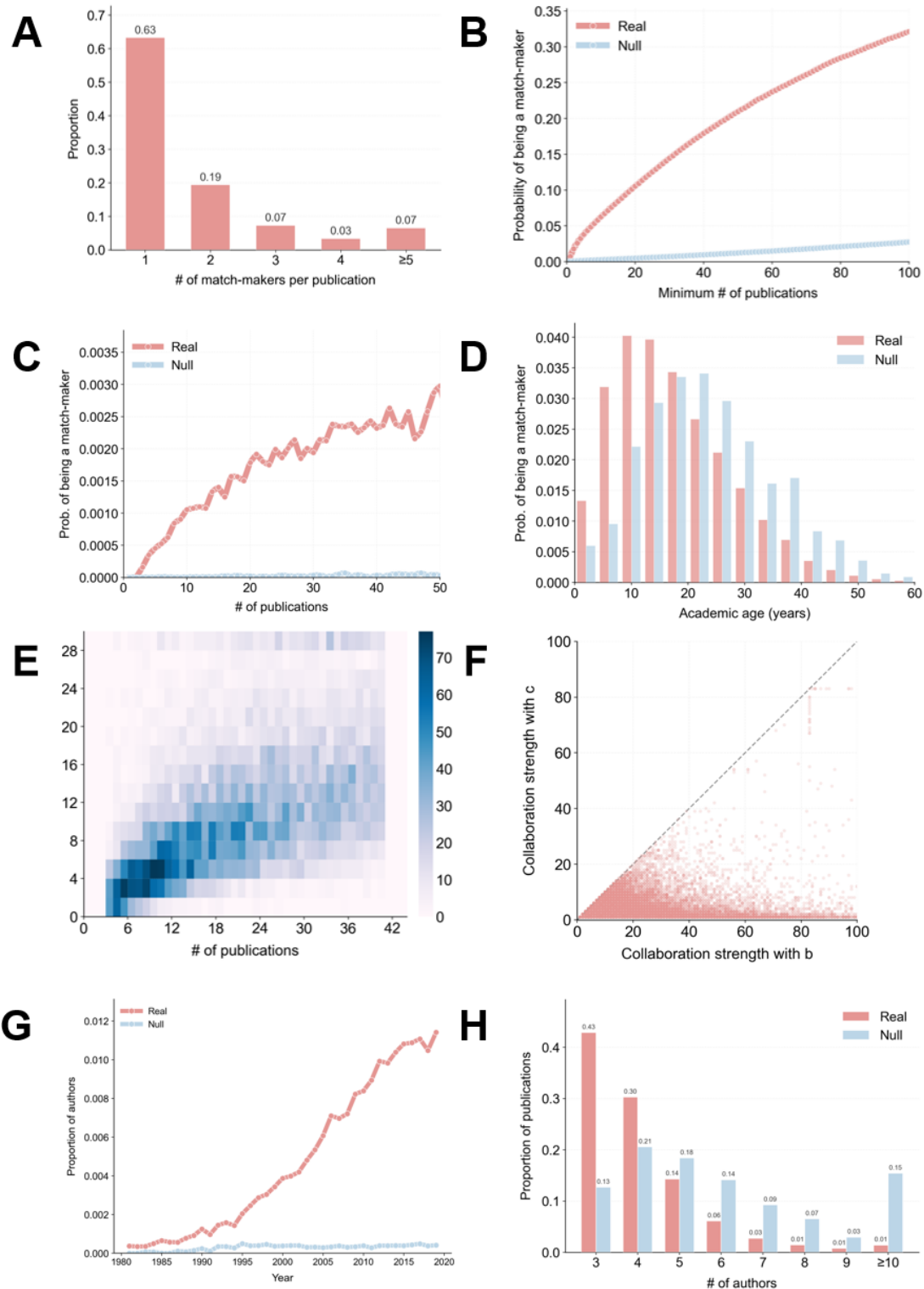

Figure S6. The findings are robust to the exclusion of b/c whose academic age was 5 years or less. (A) The distribution of the number of match-makers per publication. (B) The probability of an author acting as a match-maker, as a function of their total publication count. (C) The probability of a scientist becoming a match-maker as a function of their publication sequence. (D) The distribution of academic age when scientists first act as match-makers. (E) Relationship between publication count and academic age at the time of first match-making. (F) Collaboration frequency between the match-maker and each of the two researchers they connect. The researcher with more co-publications with the match-maker is denoted as b. (G) The annual proportion of active authors (defined as those who published at least one publication in a given year and had accumulated at least three publications by that year) who served as a match-maker from 1980 to 2019. (H) The distribution of team sizes for publications involving match-makers. The main panel is restricted to publications with only a single match-maker.

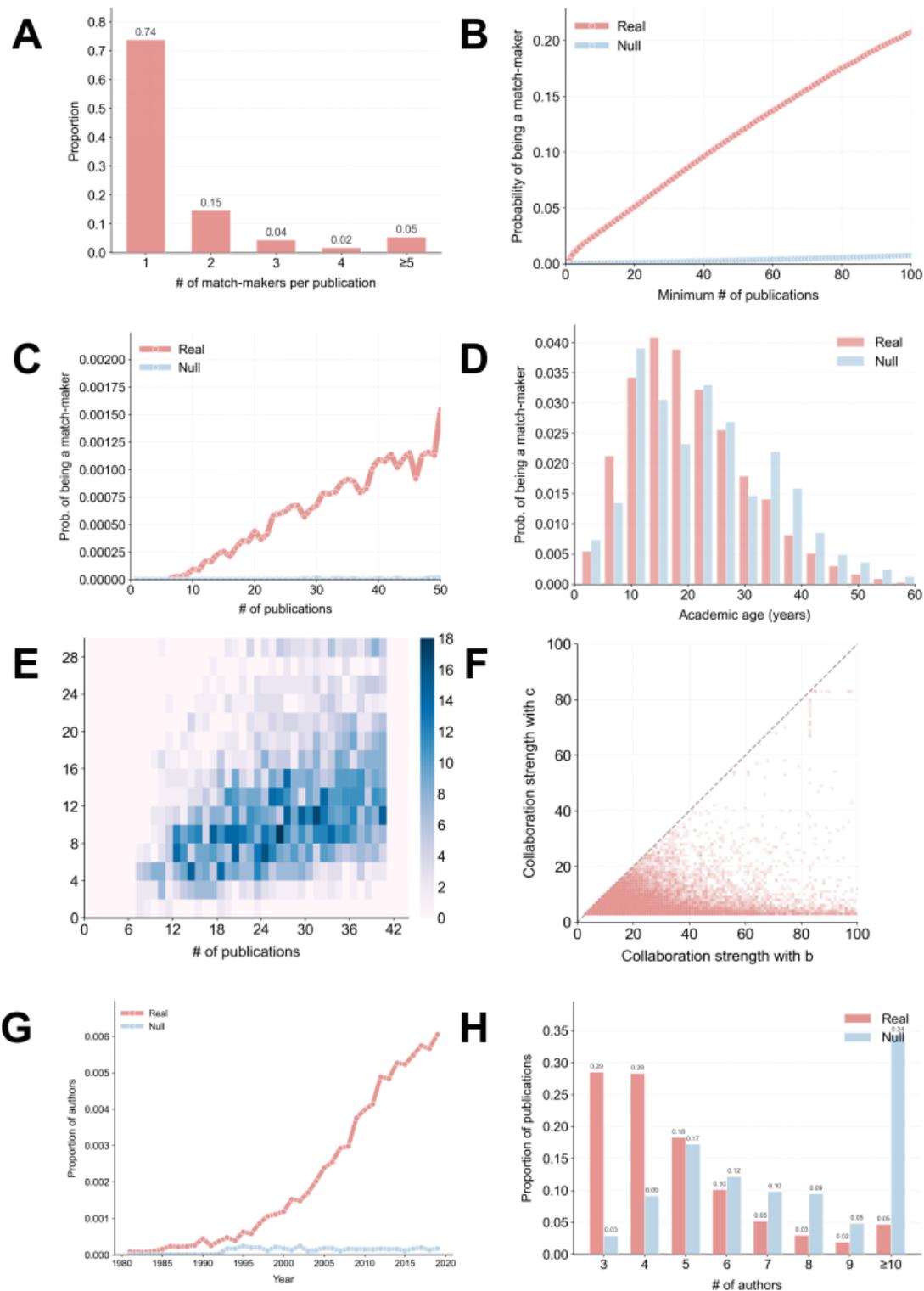

Figure S7. The findings remain robust when excluding researchers b and c with an academic age of five years or less, in combination with the requirement that the match-maker had previously co-authored at least three publications with each of the two researchers. (A) The distribution of the number of match-makers per publication. (B) The probability of an author acting as a match-maker, as a function of their total publication

count. (C) The probability of a scientist becoming a match-maker as a function of their publication sequence. (D) The distribution of academic age when scientists first act as match-makers. (E) Relationship between publication count and academic age at the time of first match-making. (F) Collaboration frequency between the match-maker and each of the two researchers they connect. The researcher with more co-publications with the match-maker is denoted as b. (G) The annual proportion of active authors (defined as those who published at least one publication in a given year and had accumulated at least three publications by that year) who served as a match-maker from 1980 to 2019. (H) The distribution of team sizes for publications involving match-makers. The main panel is restricted to publications with only a single match-maker.